\documentclass[12pt,a4paper]{article}
\usepackage{graphicx}
\makeatletter
\def\eqnarray{
\stepcounter{equation}
\let\@currentlabel=\theequation
\global\@eqnswtrue
\global\@eqcnt\z@
\tabskip\@centering
\let\\=\@eqncr
$$\halign to \displaywidth\bgroup\@eqnsel\hskip\@centering
$\displaystyle\tabskip\z@{##}$&\global\@eqcnt\@ne
\hfil$\displaystyle{{}##{}}$\hfil
&\global\@eqcnt\tw@$\displaystyle\tabskip\z@{##}$\hfil
\tabskip\@centering&\llap{##}\tabskip\z@\cr}
\makeatother

\begin{document}


\begin{titlepage}
\begin{flushright}
{KOBE-TH-01-06}\\
{DIAS-STP-01-19}\\
{TIT/HEP-472}
\end{flushright}
\vspace{5mm}
\begin{center}
{\LARGE\bf Multi-phases in gauge theories}\\
\vspace{5mm}
{\LARGE\bf on non-simply connected spaces}

\vspace{10mm}
Hisaki Hatanaka\footnote{e-mail: hatanaka@th.phys.titech.ac.jp},
Katsuhiko Ohnishi\footnote{e-mail: ohnishi@phys.sci.kobe-u.ac.jp},
Makoto Sakamoto\footnote{e-mail: sakamoto@phys.sci.kobe-u.ac.jp},
Kazunori Takenaga\footnote{e-mail: takenaga@synge.stp.dias.ie}
\vspace* {10mm} \\
{\small \it
${}^1 $Department of Physics,
Tokyo Institute of Technology, Tokyo 152-8551, Japan
\\
${}^2 $Graduate School of Science and Technology, Kobe University,
Rokkodai, Nada, \\ Kobe 657-8501, Japan
\\
${}^3 $Department of Physics, Kobe University,
Rokkodai, Nada, Kobe 657-8501, Japan
\\
${}^4 $School of Theoretical Physics,
Dublin Institute for Advanced Studies, 10 Burlington Road,
Dublin 4,  Ireland
}
\vspace{10mm}
\\

\vspace* {1cm}

\end{center}

\centerline{\bf Abstract}
\vskip0.1truein
\setlength{\baselineskip}{0.6cm}
It is pointed out that phase structures of gauge theories compactified
on non-simply connected spaces are not trivial.
As a demonstration, an $SU(2)$ gauge model on $M^3\otimes S^1$ is studied,
and it is shown to possess three phases: Hosotani, Higgs and coexisting phases.
The critical radius and the order of the phase transitions
are determined explicitly.
A general discussion about phase structures for small
and large scales of compactified spaces is given.
The appearance of phase transitions suggests
a GUT scenario in which the gauge hierarchy
problem is replaced by the dynamical problem of 
the stabilization of the radius
of a compactified space in the vicinity of
a critical radius.
\end{titlepage}

\newpage
\section{INTRODUCTION}\label{intoro}
\setlength{\baselineskip}{0.6cm}

Recently, it has been discovered that field theories
with nontrivial backgrounds on extra dimensions
have rich phase structures.
Magnetic flux passing through a circle $S^1$
can cause spontaneous breakdown
of the translational invariance of $S^1$\cite{translation}. 
A kink-like configuration
is then generated dynamically as a vacuum configuration.
A second-order phase transition
occurs at some critical radius of $S^1$,
and the translational invariance
is restored below this critical radius.
The appearance of critical radii
is one of the characteristic features of such models.
The existence of magnetic flux
influences the spectra of models
and causes nonstandard patterns
of symmetry breaking\cite{O(N) model}. \
Spontaneous breaking of translational invariance naturally
leads to a new mechanism of supersymmetry breaking\cite{SUSY},
because translations and supersymmetry transformations
are related
by the supersymmetry algebra.
\par
The magnetic flux background on one extra dimension $S^1$
can be extended to higher extra dimensions.
In Ref.\cite{monopole},
a monopole background on a sphere $S^2$
is shown to cause spontaneous breakdown
of the rotational invariance of $S^2$.
Vortex configurations are dynamically generated
as vacuum configurations,
and the number of vortices is found to be proportional
to the magnetic charge of the monopole.
A second-order phase transition occurs 
at some critical radius of $S^2$,
and the rotational symmetry turns out to be restored
below the critical radius.
\par
\begin{figure}
\includegraphics[width=15cm]{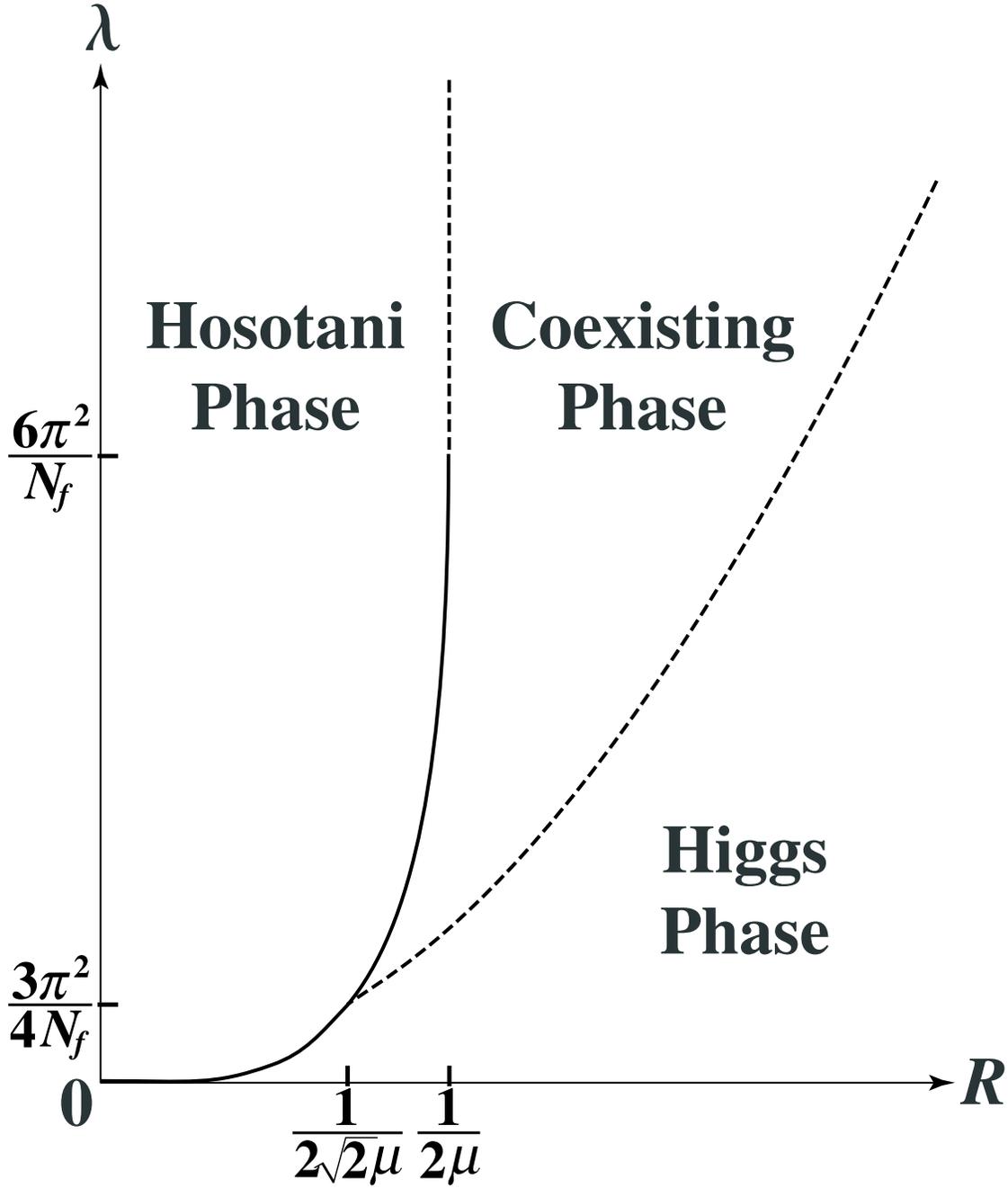}
\caption{Phase diagram of the $SU(2)$ gauge model
on $M^3\otimes S^1$.
The solid curve and dashed curves denote the first- and
second-order phase transitions, respectively.
The values $R$ and $\lambda$ are the radius of $S^1$
and the Higgs coupling, respectively.
}
\label{pdsu2}
\end{figure}

In this paper, we point out
that multi-phases can appear
in gauge theories on non-simply connected spaces
even without any background field configurations.
Studies of gauge theories,
for instance,
on $M^D\otimes S^1$
(where $M^D$ denotes a $D$-dimensional Minkowski space-time)
have a long history and have been made 
from various points of view\cite{Isham}-\cite{HIL}.
Nevertheless, as far as we know, phase structures of gauge
theories with Higgs fields
on non-simply connected spaces have not been investigated,
and a variety of phase structures of such gauge theories
have been overlooked so far.
Since the full analysis of such theories is
beyond the scope of this letter,
we restrict our consideration to a simple
$SU(2)$ gauge model
on $M^3\otimes S^1$
as a demonstration.
This model turns out to possess the three phases displayed
in Fig. \ref{pdsu2}.
The solid curve and dashed curves denote
the critical lines of the first- and second-order phase transitions,
respectively.
A nontrivial Wilson line is generated in the Hosotani phase, and
the Higgs field acquires a nonvanishing vacuum expectation value
in the Higgs phase.
The coexisting phase is a hybrid of these two phases.
\par


The paper is organized as follows.
In the next section,
an $SU(2)$ gauge model on $M^3\otimes S^1$ is introduced.
In Section \ref{phase;structure},
the phase structure of this model is clarified.
In Section \ref{gut},
as a phenomenological application of our results,
a GUT scenario is proposed,
and the manner in which the hierarchy problem is reinterpreted
in this scenario is discussed.
Section \ref{conclusion} is devoted to conclusions and discussion.
\section{$SU(2)$ Model}\label{su2model}
In order to demonstrate
that gauge theories compactified on non-simply connected spaces
are not trivial, we investigate an $SU(2)$ gauge
model on $M^3\otimes S^1$ with $N_f$ massless fermions and a Higgs boson
in the fundamental representation of $SU(2)$.
Here, $M^3$ denotes a three-dimensional Minkowski space-time, and $S^1$
is a circle of a radius $R$.
The action we consider is
\begin{equation}
S=\int d^3x\int^{2\pi R}_0dy
\bigg\{-\frac{1}{2}{\rm tr} F_{MN} F^{MN}
+\sum^{N_f}_{I=1}\bar{\psi}_I \gamma^MD_M\psi_I \nonumber \\
+(D_M\phi)^\dagger D^M\phi-V(\phi)
\bigg\},
\label{action}
\end{equation}
where $D_M=\partial_M+igA_M$ is a covariant derivative and
\begin{equation}
V(\phi)
=-\mu^2\phi^\dagger\phi+\frac{\lambda}{2}(\phi^\dagger\phi)^2.
\label{def; potential}
\end{equation}
The indices $M$ and $N$ run from $0$ to $3$,
and $x^\mu$ $(\mu=0,1,2)$ and $y$ are the coordinates
on $M^3$ and $S^1$, respectively.
All the fields are assumed to obey periodic boundary conditions
in the $S^1$ direction.\footnote{
Since $S^1$ is multiply connected,
we could impose twisted boundary conditions
on the fields,\cite{Isham}
but we are not interested in phase transitions
caused by such boundary effects\cite{translation,O(N) model,SUSY}
in this paper.
}
To investigate the vacuum configuration,
we take the vacuum expectation values of the bosonic fields
to be of the forms
\begin{eqnarray}
\langle A_\mu \rangle
&=& 0, \qquad(\mu=0,1,2),
\label{amu=0}\\
\langle A_y \rangle
&=&
\frac{1}{gR}
\left(
\begin{array}{cc}
\alpha    &    0    \\
0 & -\alpha
\end{array}
\right),
\label{ay=alpha}\\
\langle \phi \rangle
&=&
\frac{1}{\sqrt{2}}
\left(
\begin{array}{c}
v \\
0
\end{array}
\right),
\label{phi=v}
\end{eqnarray}
with real positive $v$.
\par
The leading order correction to the effective potential
for $\alpha$ comes from the fermion one-loop diagram
and is given by \cite{Hosotani}
\begin{eqnarray}
\Delta V(\alpha; R)
&=&
\frac{N_f}{2\pi^6R^4}\sum^\infty_{n=1}
\frac{\cos (2\pi n\alpha)}{n^4}
\nonumber \\
&=&
-\frac{N_f}{6\pi^2R^4}
\left\{\alpha^4-2\alpha^3+\alpha^2-\frac{1}{30}
\right\},
\label{deltav}
\end{eqnarray}
where the last equality holds only for $0\leq\alpha\leq 1$.
Although there are other one-loop corrections coming
from the gauge, ghost and Higgs fields,
we ignore them in the following analysis
to avoid unnecessary complexity.
This simplification may be justified by taking $N_f$ 
to be sufficiently large.
The effective potential for $\alpha$ and $v$
is then given by
\begin{equation}
V_{\rm eff}(\alpha, v; R)
=-\frac{\mu^2}{2} v^2+\frac{\lambda}{8} v^4+\frac{\alpha^2}{2R^2} v^2
+\Delta V(\alpha; R).
\label{def; veff}
\end{equation}
Note that the third term comes from the covariant derivative
of the Higgs field,
which gives the interaction term between the gauge and the Higgs fields.
The interaction term turns out to be crucial to determine
the phase structure of gauge theories with Higgs fields on
multiply connected spaces.
In the next section,
we determine the vacuum configuration that minimizes
the effective potential (\ref{def; veff})
and clarify the phase structure of the model.
\par
\section{Phase Structure}\label{phase;structure}
To find the vacuum configuration,
we examine the extremum conditions of the effective potential,
\begin{eqnarray}
0 &=&
\frac{\partial V_{\rm eff}}{\partial v}
=v\left[\frac{\lambda}{2} v^2-\mu^2+\frac{\alpha^2}{R^2}\right],
\label{1deri; veff; v}
\\
0
&=&
\frac{\partial V_{\rm eff}}{\partial\alpha}
=\alpha\left[\frac{v^2}{R^2}
-\frac{N_f}{3\pi^2R^4}
(2\alpha^2-3\alpha+1)\right].
\label{1deri; veff; alpha}
\end{eqnarray}
These equations lead to the following four types of solutions
as possible vacuum configurations: \\

\noindent
(I) type I
\begin{equation}
\left\{
\begin{array}{l}
\displaystyle{\alpha_{\rm I}=\frac{1}{2}}, \\
\displaystyle{\, v_{\rm I}=0},
\end{array}
\right.
\label{type1}
\end{equation}

\noindent
(II) type II
\begin{equation}
\left\{
\begin{array}{l}
\displaystyle{\alpha_{\rm II}=0}, \\
\displaystyle{\, v_{\rm II}=\sqrt{\frac{2}{\lambda}}\mu},
\end{array}
\right.
\label{type2}
\end{equation}

\noindent
(III) type III${}_{\pm}$
\begin{equation}
\left\{
\begin{array}{l}
\displaystyle{\alpha_{\rm III_{\pm}}=
\frac{3\bar{\lambda}
\pm\sqrt{4(1+2\bar{\lambda})(R\mu)^2-\bar{\lambda}(4-\bar{\lambda})}}
{2(1+2\bar{\lambda})}}, \\
\displaystyle{\, v_{\rm III{}_{\pm}}
=\sqrt{\frac{2}{\lambda}
\left(\mu^2-\frac{\alpha^2_{\rm III_{\pm}}}{R^2}\right)}},
\end{array}
\right.
\label{type3}
\end{equation}

\noindent
(IV) type IV
\begin{equation}
\left\{
\begin{array}{l}
\displaystyle{\alpha_{\rm IV}=0}, \\
\displaystyle{\, v_{\rm IV}=0},
\end{array}
\right.
\label{type4}
\end{equation}
where
\begin{equation}
\bar{\lambda}
\equiv
\frac{N_f\lambda}{6\pi^2}.
\label{lambdabar}
\end{equation}

Let us first clarify the gauge symmetry breaking for each configuration.
For the type I solution,
the gauge field for the $S^1$ direction acquires
a nonvanishing vacuum expectation value,
although it does not lead to any symmetry breaking, 
because the Wilson line
\begin{equation}
{\rm exp}
\left\{ig\int^{2\pi R}_0dy\langle A_y\rangle\right\}
={\rm exp}\left\{i\left(
\begin{array}{cc}
\pi    &    0    \\
0    &    -\pi
\end{array}\right)\right\}
=-{\bf 1}
\label{-1}
\end{equation}
is proportional to the identity matrix\cite{Hosotani}.
For the type II and III solutions,
the $SU(2)$ gauge symmetry is completely broken.
For the type IV solution,
the $SU(2)$ gauge symmetry is unbroken.
We refer to the phases of the type I,
II and III solutions as the Hosotani, Higgs
and coexisting phases, respectively.
\par
To determine which solution gives the minimum energy,
we first evaluate the effective potential for the type I, II,
IV solutions and study the stability
against small fluctuations around the type I, II, III solutions.
Since
$V_{\rm eff}(\alpha_{\rm I}, v_{\rm I}; R)
<V_{\rm eff}(\alpha_{\rm IV}, v_{\rm IV}; R)$
for any $R$,
the type IV solution is not the vacuum configuration.
Since
$V_{\rm eff}(\alpha_{\rm I}, v_{\rm I}; R)
>V_{\rm eff}(\alpha_{\rm II}, v_{\rm II}; R)$
when $R>R_1$ and 
$V_{\rm eff}(\alpha_{\rm I}, v_{\rm I}; R)
<V_{\rm eff}(\alpha_{\rm II}, v_{\rm II}; R)$
when $R<R_1$,
the type I (II) solution is not the vacuum configuration
when $R>R_1$ ($R<R_1$),
where $R_1$ is given by
\begin{equation}
R_1
=\left(\frac{\bar{\lambda}}{8}\right)^{\frac{1}{4}}\frac{1}{\mu}.
\label{def; r1}
\end{equation}
The stability arguments against small fluctuations
show that the type I (II) solution becomes unstable
when $R>R_2$ ($R<R_3$)
and that the type III${}_+$ (III${}_-$) solution is
always unstable (locally stable),
where $R_2$ and $R_3$ are given by
\begin{eqnarray}
R_2&=&\frac{1}{2\mu},
\label{def; r2}\\
R_3&=&\frac{\sqrt{\bar{\lambda}}}{\mu}.
\label{def; r3}
\end{eqnarray}
\par
It should be noted that the vacuum expectation value $\alpha$
is physically equivalent to $\alpha^\prime$
if $\alpha^\prime-\alpha\in {\bf Z}$.
This is because $\langle A_y\rangle$ itself is not a direct
physical observable, but the Wilson line
${\rm e}^{i2\pi g\langle A_y \rangle R}$
for the $S^1$ direction is.
This fact and  the symmetry of the effective potential
under $\alpha\to -\alpha$ allow us
to restrict our consideration to the range
\begin{equation}
0\leq \alpha\leq\frac{1}{2}
\label{range; alpha}
\end{equation}
without loss of generality.
Then,
for the type III${}_-$ solution, $\alpha_{\rm III_-}$,
to lie in the above region,
the radius of $S^1$ is restricted to the range
$R_4\leq R\leq R_3$ for $\bar{\lambda}\leq 1$
and $R_2\leq R\leq R_3$ for $\bar{\lambda}>1$,
where 
$R_4=\sqrt{\bar{\lambda}(4-\bar{\lambda})/4(1+2\bar{\lambda})}\mu^{-1}$.
\par
The relative magnitudes of $R_1, \cdots, R_4$ depend
on ${\bar \lambda}$.
It is not difficult to show that
$R_4 < R_3 < R_1 < R_2$ for ${\bar \lambda} < \frac{1}{8}$,
$R_4 < R_1 < R_3 < R_2$ for $\frac{1}{8} < {\bar \lambda} < \frac{1}{4}$,
$R_4 < R_1 < R_2 < R_3$ for $\frac{1}{4} < {\bar \lambda} < \frac{1}{2}$
and
$R_4 < R_2 < R_1 < R_3$ for ${\bar \lambda} > \frac{1}{2}$.%
\footnote{%
Note that $R_4$ is defined only for ${\bar \lambda} \le 4$.%
}
It turns out that the three parameter regions
i)$\bar{\lambda}<\frac{1}{8}$,
ii)$\frac{1}{8}<\bar{\lambda}<1$
and iii)$\bar{\lambda}>1$
lead to different phase structures with respect to $R$.
We study each region separately below. 

\subsection*{i) $\bar{\lambda}<\frac{1}{8}$}

It follows from the above analyses of the potential energy
and the stability that the vacuum configuration
can uniquely be determined, except in the region
$R_4<R<R_3$,
in which there are two possibilities,
type I and III${}_-$,
of the vacuum.
Comparing the effective potentials for the type I
and III${}_-$ solutions directly,
we can show that
$V_{\rm eff}(\alpha_{\rm I}, v_{\rm I}; R)
<V_{\rm eff}(\alpha_{\rm III_-}, v_{\rm III_-}; R)$
for $R_4\leq R\leq R_3$.
This fact is sufficient to determine the vacuum configuration
in the entire range of $R$.
The result is
\begin{equation}
(\alpha, v)
=\left\{
\begin{array}{ccl}
(\alpha_{\rm I}, v_{\rm I}) && \mbox{for $R<R_1$,} \\
(\alpha_{\rm II}, v_{\rm II}) &&\mbox{for $R>R_1$.}
\end{array}
\right.
\label{type1;type2}
\end{equation}
Since the type I solution is not continuously connected
to the type II solution at $R=R_1$,
a first-order phase transition occurs there. 

\subsection*{ii) $\frac{1}{8}<\bar{\lambda}<1$}

It follows from the analyses
given previously in this section
that the vacuum configuration
can uniquely be determined, except in the region
$R_4<R<R_1$ ($R_4<R<R_2$)
with $\frac{1}{8}<\bar{\lambda}<\frac{1}{2}$
($\frac{1}{2}<\bar{\lambda}<1$),
in which there are two possibilities,
the type I and III${}_-$, of the vacuum.
Comparing the effective potentials for the type I
and III${}_-$ solutions directly,
we can show that
$V_{\rm eff}(\alpha_{\rm I}, v_{\rm I}; R)
< V_{\rm eff}(\alpha_{\rm III_-}, v_{\rm III_-}; R)$
for $R_4<R<R_5$ and that
$V_{\rm eff}(\alpha_{\rm I}, v_{\rm I}; R)
> V_{\rm eff}(\alpha_{\rm III_-}, v_{\rm III_-}; R)$
for $R_5<R<R_1$ ($R_5<R<R_2$)
with $\frac{1}{8}<\bar{\lambda}<\frac{1}{2}$
($\frac{1}{2}<\bar{\lambda}<1$),
where $R_5$ is given by
\begin{equation}
R_5
=\frac{1}{2\mu}
\sqrt{\frac{3+8\bar{\lambda}-2(1+2\bar{\lambda})\sqrt{2(1+2\bar{\lambda})}
}{1+2\bar{\lambda}}}.
\label{def; r5}
\end{equation}
Thus,
we can conclude that the vacuum configuration is given by
\begin{equation}
(\alpha, v)
=\left\{
\begin{array}{ccl}
(\alpha_{\rm I}, v_{\rm I}) &&
\mbox{for $R<R_5$,} \\
(\alpha_{\rm III_-}, v_{\rm III_-}) &&
\mbox{for $R_5<R<R_3$,} \\
(\alpha_{\rm II}, v_{\rm II}) &&
\mbox{for $R>R_3$.}
\end{array}
\right.
\label{type123(ii)}
\end{equation}
Since the type I solution is not continuously connected
to the type III${}_-$
solution at $R=R_5$,
a first-order phase transition occurs there.
Since the type III${}_-$ solution becomes identical to the type II
solution at $R=R_3$,
the phase transition at $R=R_3$ is second order. 

\subsection*{iii) $\bar{\lambda}>1$}

In this case, the analyses 
given previously in this section turn out to be sufficient to
determine the vacuum configuration uniquely.
The result is
\begin{equation}
(\alpha, v)
=
\left\{
\begin{array}{ccl}
(\alpha_{\rm I}, v_{\rm I}) &&
\mbox{for $R<R_2$,} \\
(\alpha_{\rm III_-}, v_{\rm III_-}) &&
\mbox{for $R_2<R<R_3$,} \\
(\alpha_{\rm II}, v_{\rm II}) &&
\mbox{for $R>R_3$.}
\end{array}
\right.
\label{type123(iii)}
\end{equation}
Since the vacuum configuration is found to be connected
continuously at the critical radii $R=R_2$ and $R_3$,
the phase transitions are both second order.
All the results obtained above are summarized in Fig. \ref{pdsu2}.
\par
It is instructive to clarify the reason
why different phases appear for small $R$ and large $R$.
For $R\ll \mu^{-1}$,
the Higgs potential $V(\phi)$ becomes irrelevant,
and the leading contribution to the effective potential
comes from the radiative correction $\Delta V(\alpha; R)$,
so that $\alpha$ is given by the configuration that minimizes
$\Delta V(\alpha; R)$,
{\it i.e.} $\alpha=\frac{1}{2}$.
The next-to-leading term is the third term
in the effective potential (\ref{def; veff}), and
it forces $v$ to vanish with $\alpha\neq 0$.
Thus, the Hosotani phase with $\alpha=\frac{1}{2}$ and $v=0$
is expected to be realized for $R\ll\mu^{-1}$.
This is consistent with
our results obtained before.
For $R\gg \mu^{-1}$,
the radiative correction $\Delta V(\alpha; R)$
becomes irrelevant, and the leading contribution
to the effective potential
comes from the Higgs potential $V(\phi)$,
so that $v$ is given by the configuration that minimizes $V(\phi)$,
{\it i.e.} $v=\sqrt{\frac{2}{\lambda}}\mu$.
The next-to-leading term is the third term
in the effective potential (\ref{def; veff}),
and it forces $\alpha$ to vanish with $v\neq 0$.
Thus, the Higgs phase with $\alpha=0$ and $v=\sqrt{\frac{2}{\lambda}}\mu$
is expected to be realized for $R\gg\mu^{-1}$.
This is also consistent with our results obtained before.
\par
The above arguments concerning the size of $R$ can be applied
to any gauge theory with Higgs fields;
the Hosotani mechanism plays an important role
in determining the phase structure for small $R$,
while the Higgs mechanism plays an important role for large $R$.
Therefore, different symmetry breaking mechanisms
work for small and large $R$.
This is the reason for the occurrence of phase transitions.
Since the Hosotani mechanism, in general, works on gauge
theories on non-simply connected spaces,\cite{Hosotani}
the phase structure found
in the $SU(2)$ gauge model on $M^3\otimes S^1$
is expected to be a general feature of gauge theories with Higgs fields
on non-simply connected spaces.
\par
Before closing this section, we should make a few comments
on the determination of phase structures.
Strictly speaking, the argument for small (large) $R$
given above
holds only for $R \rightarrow 0$ ($R \rightarrow \infty$),
in general, 
and does not necessarily imply the existence of a Hosotani
(Higgs) phase for small (large) but {\it finite} $R$.
In fact, for instance, an $SU(4)$ gauge model with the same
matter content 
as the $SU(2)$ model has no Higgs phase\cite{nextpaper}.
Furthermore, the arguments for small and large $R$ provide no
information regarding a coexisting phase.
To determine phase structures precisely, we must study
the whole structure of effective potentials more carefully.

\section{A GUT Scenario}\label{gut}
In this section, we propose a GUT scenario
to clarify some of the phenomenological
implications of our results and discuss how the hierarchy
problem
is reinterpreted in our scenario.
\par
As an illustration, let us consider an $SU(5)$ GUT model
on $M^4\otimes S^1$, in which a Higgs field
belongs to the fundamental representation
and all mass scales are set to a GUT scale $M_G$.
\par
Suppose that the $SU(5)$ gauge symmetry is broken
to $SU(3)\times SU(2)\times U(1)$ by the Hosotani
mechanism with a nontrivial Wilson line for
small $R$
\footnote{
This symmetry breaking could be realized by choosing 
fermion matter content and/or 
the boundary conditions appropriately\cite{Hatanaka}.
}
and that the Higgs field breaks
$SU(3)\times SU(2)\times U(1)$ to
$SU(3)\times U(1)_{\rm em}$ for large $R$.
Then, a phase transition
occurs at some critical radius $R_*$ above
which $SU(2)\times U(1)$ is broken to $U(1)_{\rm em}$.
This critical
radius $R_*$ is of order $M_G^{-1}$,
which is the unique mass scale of the model.
If the radius $R$ of $S^1$ stays just above the critical radius $R_*$
and the phase transition at $R = R_*$ is second order,
the breaking scale of
$SU(5)\rightarrow SU(3)\times SU(2)\times U(1)$ is
of order $R^{-1}\sim R_*^{-1}\sim M_G$, while
the breaking scale of $SU(2)\times U(1)\rightarrow U(1)_{\rm em}$
can be much smaller than $M_G$.
This is because $SU(2)\times U(1)$ is unbroken
at $R=R_*$, and hence the breaking scale is
expected to be very small.
\par
Our GUT scenario has some advantages.
No Higgs field belonging to the adjoint representation
is necessary, because the $S^1$ component of gauge fields
plays its role.
The hierarchy problem is replaced in our scenario
by the question of why the radius of $S^1$ is so
close to the critical radius.
Although our scenario does not solve the fine-tuning problem, 
it allows us to
reinterpret the hierarchy problem as a dynamical
problem to determine the radius of $S^1$.
By minimizing the potential with respect to $R$,
we could, in principle, determine the
\lq\lq expectation value" of the radius $R$\cite{KKgravity}.
Our scenario might be stable
against quantum corrections, so that supersymmetry might not be
necessary.
It is of great interest to seek mechanisms
to stabilize the radius in
the vicinity of the critical radius.
\section{Conclusions and Discussions}\label{conclusion}
We have investigated an $SU(2)$ gauge model with a Higgs field
on $M^3\otimes S^1$
and shown that the model has three phases: Hosotani, coexisting
and Higgs phases. 
We have also obtained the critical radius and determined the
order of the phase transitions, as depicted in Fig. 1.
The non-triviality of the phase structure is not, however,
peculiar to this model but, rather, is expected to be a general feature
of gauge theories with Higgs fields on non-simply connected spaces.
Phase structures of such theories, in general, depend on matter
content as well as gauge groups.
A variety of phase diagrams appear.
Actually, if we replace the fermions in the fundamental representation
by those in the adjoint representation in our $SU(2)$ gauge model,
we obtain a phase diagram similar to Fig. \ref{pdsu2},
but in this case
the $SU(2)$ gauge symmetry is broken to $U(1)$ in the Hosotani phase.
If we replace the Higgs field in the fundamental representation
by that in the adjoint representation,
the phase diagram becomes trivial.\footnote{%
This result comes from the fact that the interaction term between
the gauge and the Higgs fields vanishes, because the vacuum
expectation values of both fields are diagonal.
}
\par
In the analysis of Section 3, we ignored contributions
from the gauge, ghost and Higgs one-loop diagrams by taking $N_f$
to be large.
For small $N_f$, they contribute to the effective potential, so that
the phase structure for small $R$ is more complicated.
A further complication arises when mass terms are added
to the fermions.
Then, quantum corrections from massive fermions with mass
$m_f$ survive for small $R$
but are exponentially suppressed for $R \gg m_f^{-1}$.
When constructing phenomenological models,
we should take them into account correctly.
\par
Other straightforward extensions of the $SU(2)$ gauge model 
consist of replacing
the $SU(2)$ gauge group by $SU(N)$ and the space-time
$M^3\otimes S^1$ by $M^D\otimes S^1$.
The latter extension will not drastically alter qualitative
features, because the $R$ dependence of radiative corrections
is not sensitive to the dimensionality of the space-time.
It turns out that the $SU(N)$ models
with $N$ even ($N\geq 4$) have
only two phases without a Higgs phase,
whereas the $SU(N)$ models with $N$ odd have
phase structures similar to that of the $SU(2)$ gauge model.
If fermions in various representations are added,
the analysis becomes more involved.
The details of these extensions will be
reported elsewhere.~\cite{nextpaper}
\par
As a phenomenological application of our discovery
of phase structures in gauge theories,
we have proposed a GUT scenario in which the gauge hierarchy problem
is replaced by the dynamical problem of the stabilization of
the radius of the compactified space in the vicinity
of a critical radius.
To construct a realistic GUT model,
it is necessary to find models with desired symmetry breaking patterns
accompanied by phase transitions
and explore the possibility of stabilizing the radius in the vicinity
of a critical radius.
This might be worth investigating.
\section*{ACKNOWLEDGMENTS}
We would like to thank J. Arafune,
K. Harada, Y. Hosotani, T. Inagaki,
K. Inoue, T. Kashiwa, C. S. Lim,
K. Ogure, N. Sakai, K. Shiraishi, M. Tachibana,
S. Tanimura, H. Terao
and T. Tsukioka
for valuable discussions and useful comments.
This work was supported in part by a
JSPS Research Fellowship for Young Scientists
(H.H.) and by a Grant-In-Aid for Scientific Research
(No.12640275) from the Ministry of Education,
Science and Culture,
Japan (M.S.).
K.T.  would like to thank the Dublin Institute for
Advanced Studies for warm hospitality.


\begin{thebibliography}{99}
\bibitem{translation}
M. Sakamoto, M. Tachibana, K. Takenaga,
Phys. Lett.
{\bf B457} (1999), 33, hep-th/9902069.
\bibitem{O(N) model}
K. Ohnishi, M. Sakamoto,
Phys. Lett. {\bf B486} (2000), 179, hep-th/0005017. \\
H. Hatanaka, S. Matsumoto, K. Ohnishi, M. Sakamoto,
Phys. Rev. {\bf D63} (2001), 105003, hep-th/0010283;
in {\it Proceedings of the 30th International Conference
on High Energy Physics, 2000}, edited by C. S. Lim
and T. Yamanaka
(World Scientific, Singapore, 2001),
hep-th/0010041.
\bibitem{SUSY}
M. Sakamoto, M. Tachibana, K. Takenaga,
Phys. Lett. {\bf B458} (1999), 231, hep-th/9902070;
Prog. Theor. Phys. {\bf 104} (2000), 633, hep-th/9912229;
in {\it Proceedings of the 30th International Conference
on High Energy Physics, 2000}, edited by C. S. Lim
and T. Yamanaka
(World Scientific, Singapore, 2001),
hep-th/0011058.
\bibitem{monopole}
S. Matsumoto, M. Sakamoto, S. Tanimura,
Phys. Lett. {\bf B518} (2001), 163,
hep-th/0105196. \\
M. Sakamoto, S. Tanimura, hep-th/0108208,
Phys. Rev. {\bf D65} (2002) 065004.
\bibitem{Isham}
{C. J. Isham,
Proc. R. Soc. London {\bf A362} (1978), 383;
{\bf A364} (1978), 591.
}
\bibitem{inducedmass}
L.H.  Ford, T. Yoshimura,
Phys. Lett. {\bf A70} (1979), 89. \\
D.J.  Toms,
Phys. Rev. {\bf D21} (1980), 928;
{\bf D21} (1980), 2805. \\
G. Denardo, E. Spallucci,
Nucl. Phys. {\bf B169} (1980), 514;
Nuovo Cim.
{\bf A58} (1980), 243.
\bibitem{Ssmechanism}
J. Scherk, J.H.  Schwarz, Phys. Lett. {\bf B82} (1979), 60. \\
P. Fayet, Phys. Lett. {\bf B159} (1985), 121;
 Nucl. Phys. {\bf B263} (1986), 649. \\
K. Takenaga, Phys. Lett. {\bf B425} (1998), 114,
hep-th/9710058; Phys. Rev. {\bf D58} (1998), 026004;
{\bf D61} (2000), 129902(E), hep-th/9801075;
Phys. Rev. {\bf D64} (2001), 066001, hep-th/0105053.
\bibitem{Hosotani}
Y. Hosotani, Phys. Lett. {\bf B126} (1983), 309;
Ann. Phys. {\bf 190} (1989), 233.
\bibitem{HP}
A. Higuchi, L. Parker,
Phys. Rev.
{\bf D37} (1988), 2853.
\bibitem{HIL}
{H. Hatanaka, T. Inami, C.S.  Lim,
Mod. Phys. Lett.  {\bf A13} (1998), 2601, hep-th/9805067.
}
\bibitem{Hatanaka}
H. Hatanaka,
Prog. Theor. Phys. {\bf 102}
(1999), 407, hep-th/9905100. \\
C.L. Ho, Y. Hosotani,
Nucl. Phys. {\bf B345} (1990), 445.
\bibitem{KKgravity}
T. Appelquist, A. Chodos,
Phys. Rev. {\bf D28}
(1983), 772.
\bibitem{nextpaper}
H. Hatanaka, K. Ohnishi, M. Sakamoto, K. Takenaga,
in preparation.
%
%
%
%
%
\end{thebibliography}
\end{document}